\documentclass[aps,twocolumn,pra,showpacs]{revtex4}
\usepackage{amssymb}
\usepackage{mathrsfs}
\usepackage{graphicx}
\usepackage{amsmath,amssymb,amsthm,bbm,latexsym}
\usepackage{amsthm}
\usepackage{amsbsy}
\usepackage{epsfig}
\usepackage{graphicx}
\usepackage{float}
\usepackage{dcolumn}
\usepackage{amsmath}

\begin{document}

\author{Zhao-Ming Wang$^{1}$\footnote{%
Email address: mingmoon78@126.com}, C. Allen Bishop$^{2}$, 
Yong-Jian Gu$^{1}$\footnote{%
Email address: yjgu@ouc.edu.cn}, and Bin Shao$%
^{3}$}
 \affiliation{$^{1}$ Department of
Physics, Ocean University of China, Qingdao, 266100, China}
\affiliation{$^{2}$  Department of Physics, Southern Illinois
University, Carbondale, Illinois 62901-4401}
\affiliation{$^{3}$
 Department of Physics, Beijing Institute of Technology, Beijing,
100081, China}

\title{Duplex quantum communication through a spin chain}

\begin{abstract}

Data multiplexing within a quantum computer can allow for the simultaneous 
transfer of multiple streams of information over a shared medium 
thereby minimizing the number of channels needed for requisite data transmission. 
Here, we investigate a two-way quantum communication protocol using a spin
chain placed in an external magnetic field. In our scheme, Alice and
Bob each play the role of a sender and a receiver as 
two states $\cos (\frac{\theta
_{1}}{2})\left\vert 0\right\rangle+\sin (\frac{\theta
_{1}}{2})e^{i\phi _{1}}\left\vert 1\right\rangle$, $\cos
(\frac{\theta _{2}}{2})\left\vert 0\right\rangle+\sin (\frac{\theta
_{2}}{2})e^{i\phi _{2}}\left\vert 1\right\rangle$ are transferred
through one channel simultaneously.  We find that the transmission fidelity 
at each end of a spin chain can usually be enhanced by the presence of a 
second party. This is an important result for establishing the viability of duplex 
quantum communication through spin chain networks. 
\end{abstract}

\maketitle

\section{Introduction}

In classical electronic communications, full-duplex transmission capabilities
allow for the simultaneous sending and receiving of data to and from some 
remote host or process. In instances where real-time information transfer
is required between two parties, for example in voice communications and
high-performance distributed computing, full-duplex transmission is desired \cite{tan}. 
While two physical twisted-pairs of wires per cable provide the avenue for
duplex transmission classically, there has been no counterpart for such communications in quantum machines. It will be shown, however, that spin chains can be
used as foundational elements of quantum duplex communications. In principal, full-duplex information transfer can be achieved during quantum 
computations using the interactions which naturally occur between 
neighboring sites of a spin chain. 

It was Bose 
who first suggested 
using an unmodulated spin chain to serve as 
a mediator for quantum information 
transfer \cite{Bose2003}. The basic idea goes like this: An arbitrary qubit 
state is encoded at one end of the chain which then evolves naturally 
under spin
dynamics. Later, at some time $\tau$, the state can be received 
at the other end with some probability. Although his original proposal 
offers the advantage of simplicity, it does not allow for a perfect 
state transfer in most situations. In order to improve the fidelity of 
transmission an extensive investigation has been made regarding 
state or entanglement transfer through
permanently coupled spin chains \cite{Bose2003,Giovanetti2006,Gualdi2008,Wang2007,Wang2009,Allen2010,Durgarth2005,Durgarth2007}.
The transmission fidelity (entanglement) can be
significantly enhanced by means of introducing phase shifts, energy
currents \cite{Wang2007}, or by properly encoding the state over 
more than one site \cite{Wang2009,Allen2010}. 
There are also methods using two parallel spin chains which allow for a
perfect state transfer (PST) \cite{Durgarth2005,Durgarth2007}. In this
case PST is achieved using measurements at the end
of the chain. Other methods require a single local on-off switch
actuator \cite{Schirmer2009}, a single-spin optimal
control \cite{XiaotingWang2010}, or via certain classes of random
unpolarized  spin chains \cite{Yao2011}. PST in a strongly coupled
antiferromagnetic spin chain has been reported in Ref.~\cite{Oh2011} which
requires weakly coupled external qubits. Furthermore, PST \cite{Wu20091} or perfect function
transfer \cite{Wu20092} can also be realized in a variety of interacting media, including, but not 
limited to, the spin chain model.

In most scenarios considered in the literature the communication is assumed to occur 
in one direction, i.e. if Alice sends the information Bob plays the role of the receiver. 
This type of communication resembles a form of broadcasting where one party sends a signal while the second party simply 
``listens''. Although broadcasting quantum information is certainly an important 
method of communication, it is by no means the only method needed 
for quantum computation. Full scale quantum computing will undoubtedly require multiplexing 
between multiple processes and therefore an analysis of the effects of state transmission in 
two directions is warranted. Here we study a duplex quantum communication protocol using a spin chain 
placed in a external magnetic field. In this case 
Alice and Bob each play the role of a sender and a receiver. Unlike the current trend in spin 
chain research, our intention is not to improve the quality of state transfer 
but rather to investigate how the presence of a second party effects 
the other senders transmission. We focus on the least technically 
challenging spin chain model and simply require 
local control over the first and last site for the preparation and reception of the
states. We find that in most cases the presence of a second party can significantly enhance the 
fidelity of state transmission from the other party thereby allowing for reliable two-way communication.

The paper will be outlined as follows. In Section II we will describe the 
physical model we consider and derive expressions for the communication fidelity. The 
numerical results obtained from these expressions will be presented in Section III. Finally, 
we will conclude with a summary of our findings in Section IV.

\section{The model}

We depict our scheme in Fig.~1. Alice and Bob are situated at opposite
ends of a one-dimensional array of $N$ spin-1/2 systems. We assume 
the chain has been cooled to the ground state 
$\left\vert \downarrow_1 \downarrow_2 \hdots \downarrow_N  \right\rangle$
prior to the 
encoding process, where we have defined the eigenstates of the Pauli 
operator $\sigma_z$ to be   
$\left\vert \downarrow \right\rangle \equiv \left\vert 0 \right\rangle$ and 
$\left\vert \uparrow \right\rangle \equiv \left\vert 1 \right\rangle$. 
Alice and Bob then respectively prepare the states $\left\vert \varphi _{1}\right\rangle $ and $\left\vert
\varphi _{2}\right\rangle $ which they intend to send. To simplify matters, 
we will assume that both encodings take place simultaneously.
\begin{figure}[tbph]
\centering
\includegraphics[scale=1.2,angle=0]{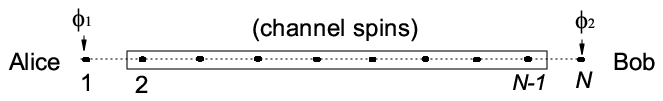}
\par
\caption{Schematic illustration of our communication protocol.
Alice and Bob respectively encode the
states $\left\vert \varphi _{1}\right\rangle$, $\left\vert \varphi
_{2}\right\rangle$ into the spins located at the first and last sites 
of the chain. After some time $\tau$, they attempt to receive the state which 
was sent from the opposite end.}
\par
\label{fig:1}
\end{figure}
\begin{figure}[tbph]
\centering
\includegraphics[scale=0.6,angle=0]{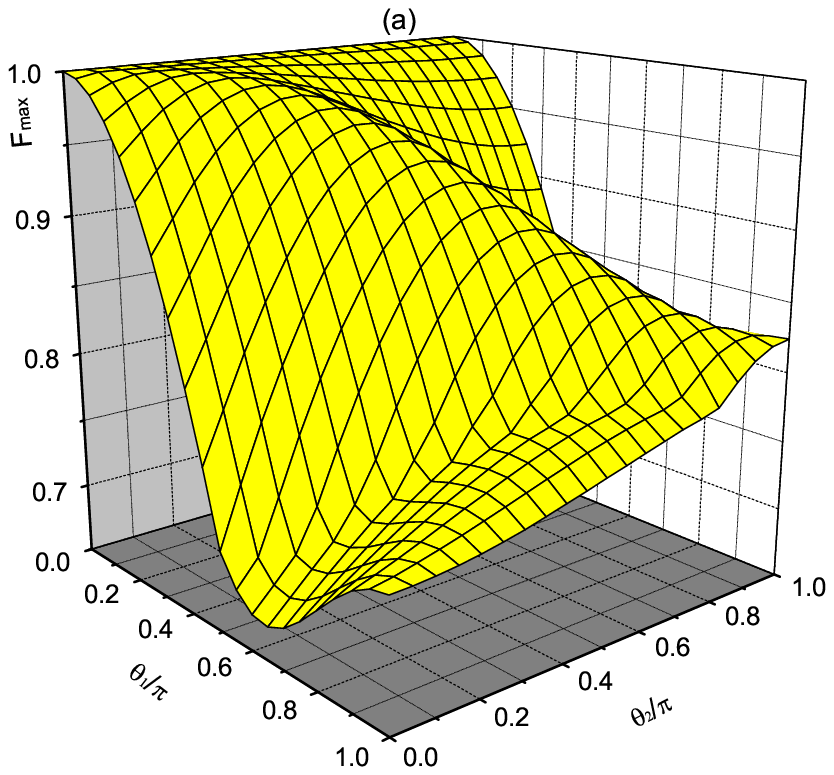} %
\includegraphics[scale=0.6,angle=0]{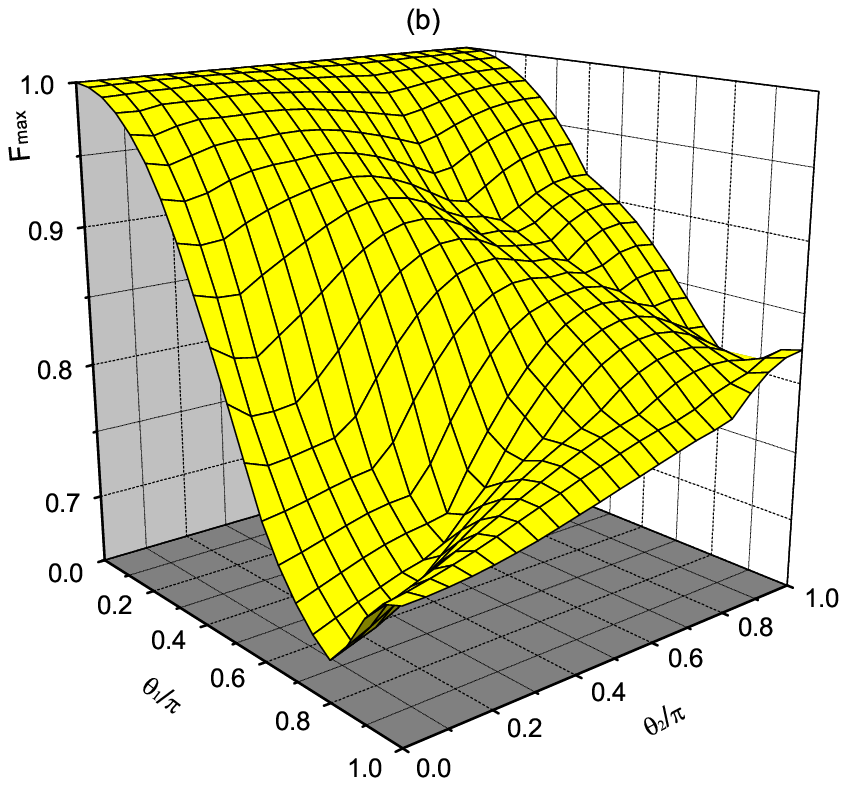} %
\includegraphics[scale=0.6,angle=0]{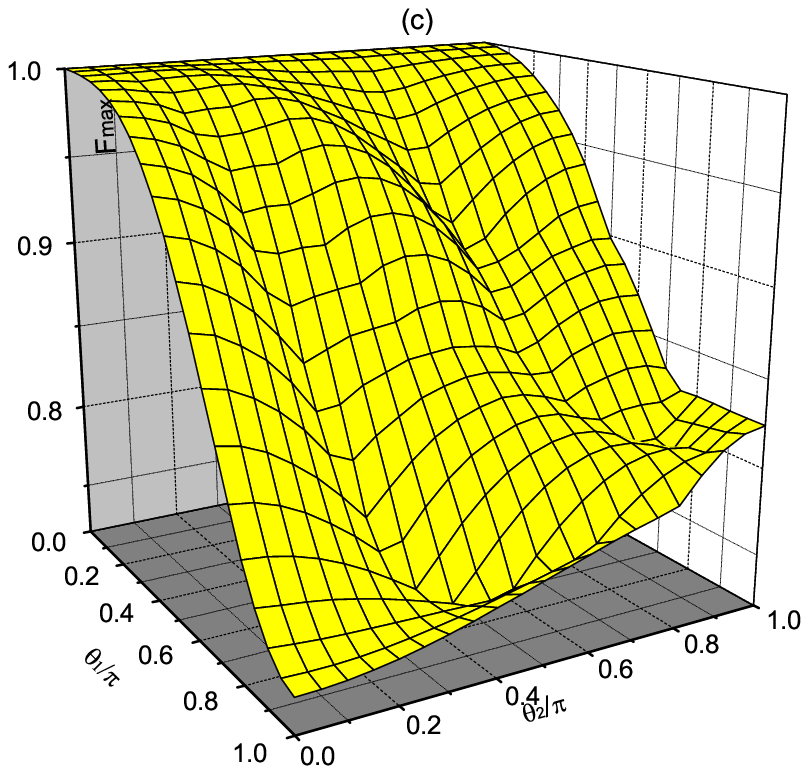}
\par
\caption{(Color online.) The maximum fidelity $F_{max}$ 
which can be obtained when 
transferring a quantum state from Alice to Bob in a time interval
t $\in[10,50]$ as a function of the polar angles $\theta _{1}$ and
$\theta _{2}$. We have set $N=10$, $\phi _{1}=\phi _{2}=0$ and (a) $h=0.0$
(b) $h=0.1$ and (c) $h=1.0$.}
\par
\label{fig:2}
\end{figure}
After the states of the spin systems at sites 1 and $N$ have been prepared
the system as a whole will then be allowed to evolve. This evolution 
will be generated by 
nearest-neighbor XY-type interactions and an externally applied 
magnetic field
\begin{equation}
H=-\frac{J}{2}\sum\limits_{i=1}^{N-1}(\sigma _{i}^{x}\sigma
_{i+1}^{x}+\sigma _{i}^{y}\sigma _{i+1}^{y})-h\sum\limits_{i=1}^{N}\sigma
_{i}^{z}.
\end{equation}%
We assume a ferromagnetic coupling and take the interaction constant 
to be $J=1.0$
throughout. The constant $h$ represents the external magnetic field strength 
of a field applied along the $z$ direction and $%
\sigma _{i}^{x,y,z}$ denote the Pauli operators acting on spin $i$. We
consider an open ended chain which is perhaps the most natural geometry for a
channel. This Hamiltonian can be 
diagonalized by means of the Jordan-Wigner
transformation that maps spins to one-dimensional spinless fermions with
creation operators defined by $c_{l}^{\dag }=(\prod\limits_{s=1}^{l-1}-\sigma
_{l}^{z})\sigma _{l}^{+}$. Here $\sigma _{l}^{+}=\frac{1}{2}$ $(\sigma
_{l}^{x}+i\sigma _{l}^{y})$ denotes the spin raising operator at site $l$.
The action of $c_{l}^{\dag }$ is to flip the spin at site $l$ from down to
up. For indices $l$ and $m$, the operators $c_{l}$ and $c_{m}^{\dag }$ satisfy 
the anticommutation relations $%
\{c_{l},c_{m}^{\dag }\}=\delta _{lm}$. The z-component of the total spin is
a conserved quantity implying the conservation of the total number of
excitations $M = \sum_{l}c_{l}^{\dag }c_{l}$ in the system. The evolution
of the creation operator $c_{j}^{\dag }$ is given by \cite{Amico}

\begin{equation}
c_{j}^{\dag }(t)=\sum_{l}f_{j,l}(t)c_{l}^{\dag },
\end{equation}%
where
\begin{equation}
f_{j,l}=\frac{2}{N+1}\sum\limits_{m=1}^{N}\sin (q_{m}j)\sin
(q_{m}l)e^{-iE_{m}t},
\end{equation}%
with $E_{m}=2h-2J\cos (q_{m})$ and $q_{m}=\pi m/(N+1).$ When the number of magnon
excitations is more than one, the time evolution of the creation operators is %
\cite{Wichterich}
\begin{equation}
\label{Eq:multiple}
\prod\limits_{m=1}^{M}c_{j_{m}}^{\dag }(t)=  \!\!\!\!\!\!\! 
\sum\limits_{l_{1}<...<l_{M}} \!\!\!\!\!   \det
\left\vert
\begin{array}{cccc}
f_{j_{1},l_{1}} & f_{j_{1},l_{2}} & ... & f_{j_{1},l_{M}} \\
f_{j_{2},l_{1}} & f_{j_{2},l_{2}} & ... & f_{j_{2},l_{M}} \\
... & ... & ... & ... \\
f_{j_{M},l_{1}} & f_{j_{M},l_{2}} & ... & f_{j_{M},l_{M}}%
\end{array}%
\right\vert \prod\limits_{m=1}^{M}c_{l_{m}}^{\dag }
\end{equation}%
where $M$ gives the number of excitations. The set \{$j_{1},j_{2},...,j_{M}$%
\} denotes the sites where the excitations are created and \{$%
l_{1},l_{2},...,l_{M}$\} denotes an ordered combination of $M$ different
indices from $\{1,2,...,N\}$. In this paper we consider 
chains which carry no more than two excitations, so we set $M=2$ 
in the situations when Eq.~(\ref{Eq:multiple}) is used.

To proceed, let us assume that Alice prepares an arbitrary qubit 
state $\left\vert \varphi _{1}\right\rangle =$
$\alpha _{1}\left\vert 0\right\rangle +\beta _{1}\left\vert 1\right\rangle $
at the first site while Bob prepares the state $\left\vert \varphi _{2}\right\rangle =$ $\alpha
_{2}\left\vert 0\right\rangle +\beta _{2}\left\vert 1\right\rangle $ at the 
$N$th site. The
initial state of the chain is then
\begin{equation}
\label{Eq:initial}
\left\vert \Phi (t=0)\right\rangle =(\alpha _{1}\left\vert 0\right\rangle
+\beta _{1}\left\vert 1\right\rangle )\otimes \left\vert \mathbf{0}%
\right\rangle \otimes 
(\alpha _{2}\left\vert 0\right\rangle +\beta
_{2}\left\vert 1\right\rangle ).
\end{equation}%
The time evolution of $\left\vert \Phi (0)\right\rangle $ will be
\begin{equation}
\left\vert \Phi (t)\right\rangle =[\alpha _{1}\alpha
_{2}+\sum\limits_{i=1}^{N}A_{i}(t)c_{i}^{\dag
}+\sum\limits_{i<i^{\prime}}B_{i,i^{\prime }}(t)c_{i}^{\dag }c_{i^{\prime
}}^{\dag }]\left\vert \mathbf{0}\right\rangle,  \label{eq:evo}
\end{equation}%
where $A_{i}=\alpha _{1}\beta _{2}f_{N,i}+\beta _{1}\alpha _{2}f_{1,i}$, $%
B_{i,i^{\prime }}=\beta _{1}\beta _{2}(f_{1,i}f_{N,i^{\prime
}}-f_{1,i^{\prime }}f_{N,i})$. Although we have used 
$\left\vert \mathbf{0}\right\rangle$ interchangeably in these last 
two equations, it should be understood that in Eq.~(\ref{Eq:initial}) 
the notation $\left\vert \mathbf{0}\right\rangle$ is used to refer 
to the state 
$\left\vert \downarrow_2 \downarrow_3 \hdots \downarrow_{N-1} \right\rangle$
of the channel spins between the first and last site, while in 
Eq.~(\ref{eq:evo}) it is used to represent the state 
$\left\vert \downarrow_1 \downarrow_1 \hdots \downarrow_{N} \right\rangle$ 
of the entire chain. We see from Eq.~(\ref{eq:evo}) that the
excitations which are created at the ends of the chain begin to 
spread over time, resulting in a probability distribution of appearing 
over all sites. Unlike the original one way communication
protocol where at most one excitation exists in the chain, here we can 
find excitations at sites $i$ and $i^{\prime}$ ($i \neq i^{\prime}$) 
with probability $\left\vert B_{i,i^{\prime }}(t)\right\vert ^{2}$.
Clearly, when $\alpha _{2}=1.0$ and $\beta _{2}=0.0,$ our two-way 
communication scheme reduces to the original scenario \cite{Bose2003}.

Now consider the dynamics of the system. The fidelity between the
received state and initial state $\left\vert \varphi_i\right\rangle $ 
$(i=1,2)$ is defined by $F=\sqrt{\left\langle \varphi_i\right\vert \rho (t)\left\vert \varphi_i\right\rangle },$ where
$\rho (t)$ is the reduced density matrix associated with 
the spin state at the receiving position. 
In what follows we will assume that both parties measure 
their states simultaneously since the measurement process 
will necessarily disturb the 
state of the system. It would be interesting 
to examine the effect of non-simultaneous measurements on state transmission and will be 
a subject of later work. For now, with the assumption of simultaneous measurements, 
the fidelity at Bob's end $F_{N}$ and Alice's end $F_{1}$ are calculated to be
\begin{widetext}
\begin{equation}
F_{N}=\left[\left\vert \left\vert \alpha _{1}\right\vert ^{2}\alpha _{2}+\beta _{1}^{\ast }A_{N}\right\vert ^{2}+\sum\limits_{i=1}^{N-1}\left\vert \alpha
_{1}^{\ast }A_{i}+\beta _{1}^{\ast }B_{i,N}\right\vert ^{2}+\sum\limits_{i<(i^{\prime }\neq N)}\left\vert \alpha _{1}\right\vert ^{2}\left\vert
B_{i,i^{\prime }}\right\vert ^{2}\right]^{1/2}
\end{equation}
\begin{equation} \label{eq:alice}
F_{1}=\left[\left\vert \left\vert \alpha _{2}\right\vert ^{2}\alpha _{1}+\beta _{2}^{\ast }A_{1}\right\vert ^{2}+\sum\limits_{i=2}^{N}\left\vert \alpha
_{2}^{\ast }A_{i}+\beta _{2}^{\ast }B_{1,i}\right\vert ^{2}+\sum\limits_{(i\neq 1)<i^{\prime }}\left\vert \alpha _{2}\right\vert ^{2}\left\vert
B_{i,i^{\prime }}\right\vert ^{2}\right]^{1/2}
\end{equation}
\end{widetext}
We will analyze the behavior of duplex quantum communication in terms 
of these fidelity measures next. We will show that the transmission fidelity 
at each end can be significantly enhanced by the presence of the other party.

\section{Results and discussion}

We assume that Alice and Bob respectively prepare the arbitrary qubit 
states $\alpha
_{1}\left\vert 0\right\rangle +\beta _{1}\left\vert 1\right\rangle $ and $%
\alpha _{2}\left\vert 0\right\rangle +\beta _{2}\left\vert 1\right\rangle$
with each individual state being represented by a 
point on a Bloch sphere with $\alpha _{i}=\cos (\frac{\theta
_{i}}{2})$, and $\beta _{i}=\sin (\frac{\theta _{i}}{2})e^{i\phi
_{i}}(i=1,2)$. First we investigate the effect of the polar angles
$\theta_i $ on the transmission fidelity and let
$\phi _{1}=\phi _{2}=0$. In Fig. 2, we plot the maximal fidelity which 
can be obtained at
Bob's side of a $N=10$ site chain in a time interval $t\in \lbrack
10,50]$ as a function of the parameters $\theta _{1}$ and $\theta _{2}$. 
We exclude the time interval $[0,10]$ 
since we need to avoid the fact that if Alice and
Bob both send a similar state, the fidelity will be very high at
$t=0$ regardless of any actual state transfer. Throughout
the paper, the maximum fidelity $F_{\max }$ is found in the
same time interval. When the chain is isolated from an external field 
($h=0$) we find that when $\theta_{1}=\theta_{2}$, i.e. when the two
states are identically prepared, $F_{\max}$ is higher than in the other cases 
(see Fig. 2(a)). Although the fidelity is maximized when both 
senders encode similar states, any nonzero $%
\theta _{2}$ will enhance the fidelity of Alice's transmission when 
$h=0$ and $\phi _{1}=\phi _{2}=0$. 
For example, when $%
\theta _{1}/\pi=0.6$, $F_{\max }=0.67$ for $\theta _{2}=0,$ while $F_{\max
}=0.91$ for $\theta _{2}/\pi=0.6$. Since the chain has 
been initialized to the ground state $\left\vert \downarrow_1 \downarrow_2 \hdots \downarrow_N  \right\rangle$ before Alice and Bob encode their 
states, a $\theta _{2}=0$ encoding at Bob's end amounts to the 
same thing as if he were not even present. We can therefore 
infer that duplex quantum communication has a positive 
impact on a senders ability to transfer information, at least in the 
case where $h=0$ and $\phi _{1}=\phi _{2}=0$. In a way, 
Bob's encoding resembles an "activator" used in chemistry.

Let us now consider the influence of the external magnetic field. 
Fig. 2(b) and (c)
exemplify the weak field $(h=0.1)$ and strong field ($h=1.0$) regimes. In
Fig. 2(b), the behavior of $F_{\max}$ with $\theta_{1}$ and
$\theta_{2}$ is similar to that given in Fig 2(a), but Fig 2(c) shows 
that the presence of a strong
field can hinder the aforementioned properties as $F_{\max}$ is only
slightly enhanced for some range of $\theta _{2}$. For some values of
$\theta _{2}$, the fidelity of Alice's transmission can actually 
decrease, though the decrease is small. For
example, when $\theta _{1}/\pi=0.8$, $F_{\max }=0.79$ for 
$\theta_{2}=0$ while $F_{\max }=0.75$ for $\theta _{2}/\pi=0.35$. 
The fidelity of Bob's transmission can be explicitly calculated 
from Eq.~(\ref{eq:alice}). The results will be similar to those above due to 
symmetry hence we only consider Alice's state transfer here. 
\begin{figure}[tbph]
\centering
\includegraphics[scale=0.6,angle=0]{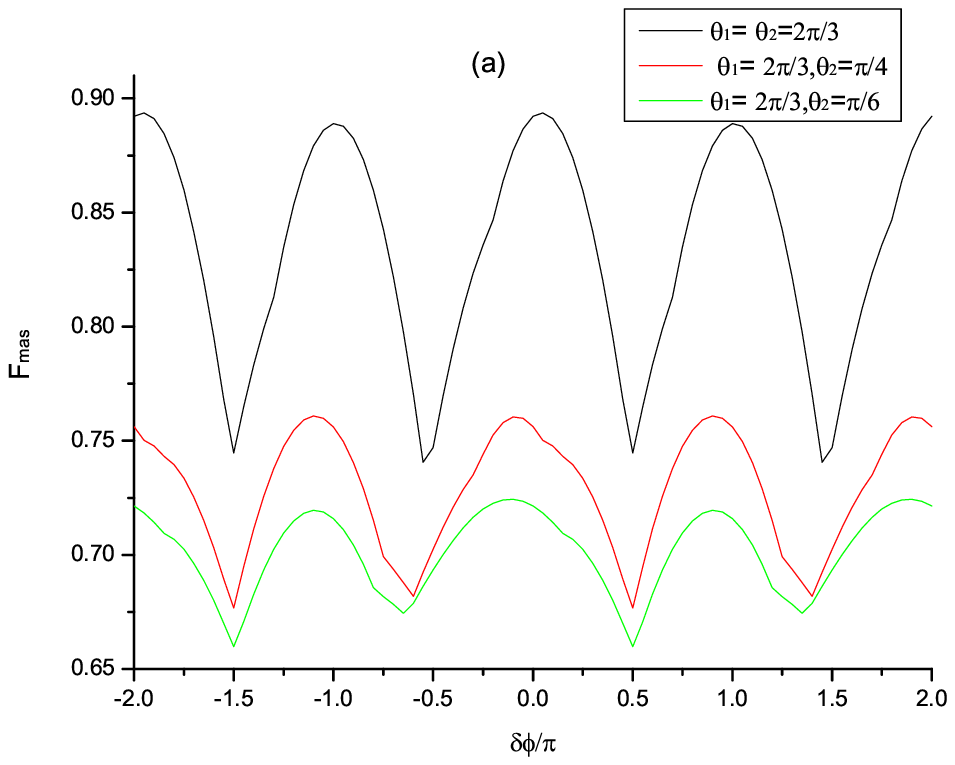} %
\includegraphics[scale=0.6,angle=0]{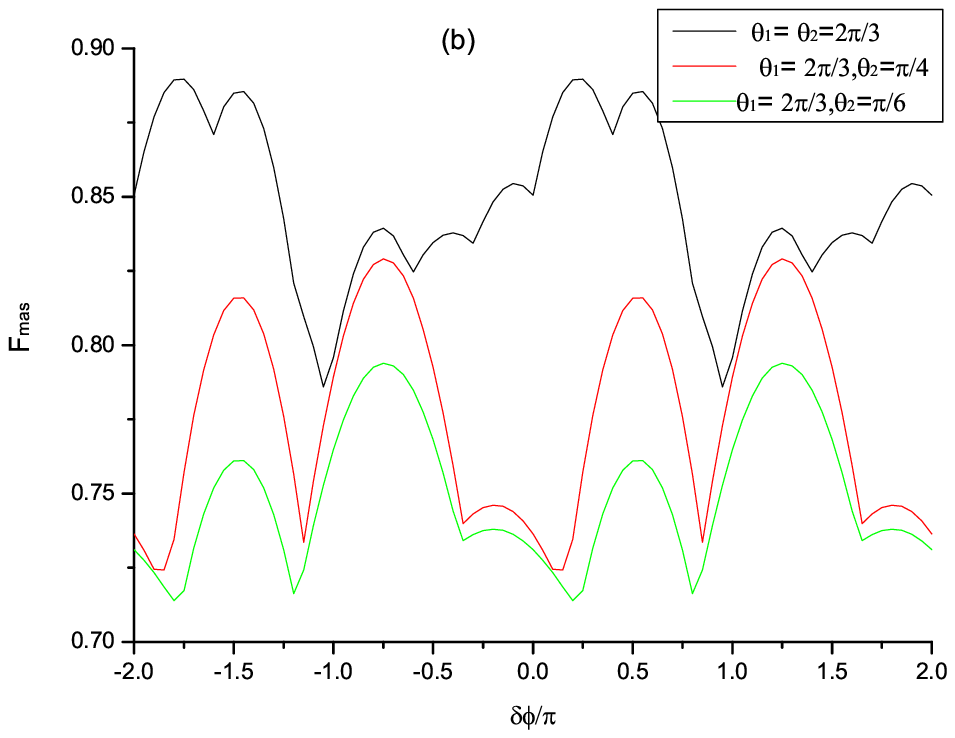} %
\includegraphics[scale=0.6,angle=0]{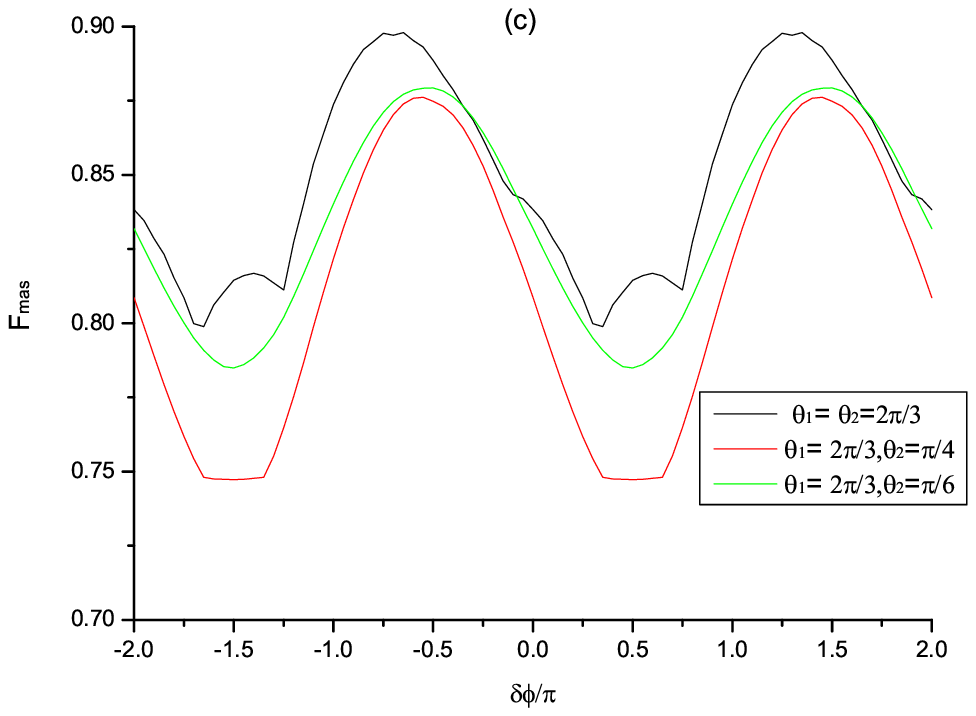}
\par
\caption{(Color online.) The maximum fidelity $F_{max}$ of Alice's state 
transfer as a function of $\delta\phi= \phi_2- \phi_1$. We search 
for a maximum in the time interval t$\in[10,50]$ for 
fixed parameters $\theta_i$. Here we have 
set $N=10$, (a) $h=0.0$, (b) $h=0.1$, and (c) $h=1.0$.}
\par
\label{fig:3}
\end{figure}
We now consider the effect of the phase angles $\phi_i$. A numerical calculation shows that
$F_{\max }$ only depends on the difference of the parameters $\phi
_{1}$, and $\phi _{2}$. In Fig.~3, we plot the maximum fidelity $F_{\max }$ as a function of the difference $\delta \phi = \phi_2 -\phi_1$ 
for fixed parameters $\theta_{1}$ and $\theta_{2}$. Again, we consider the 
vanishing field ($h=0.0$), weak field ($h=0.1$), and strong field ($h=1.0$) regimes. 
We find that for an isolated chain ($h=0.0$) the fidelity of Alice's state transfer 
will be greatest when the difference $\delta \phi \approx k\pi$ for $(k=-2,-1,0,1,2)$. When 
the difference $\delta \phi/ \pi \approx -1.5,-0.5,0.5,1.5$ the maximum fidelity which can be obtained 
in the time interval t$\in[10,50]$ will minimized with respect to $\delta \phi$. 
When an external field is applied to the chain it is more difficult 
to assess the behavior of state transfer as can be seen in 
Fig. 3(b) and (c). However, in all three cases we find that if Alice and 
Bob both choose $\theta _{1}=\theta _{2}$ the fidelity will be greater when 
compared to other values of the parameters $\theta_i$.
\begin{figure}[tbph]
\centering
\includegraphics[scale=0.6,angle=0]{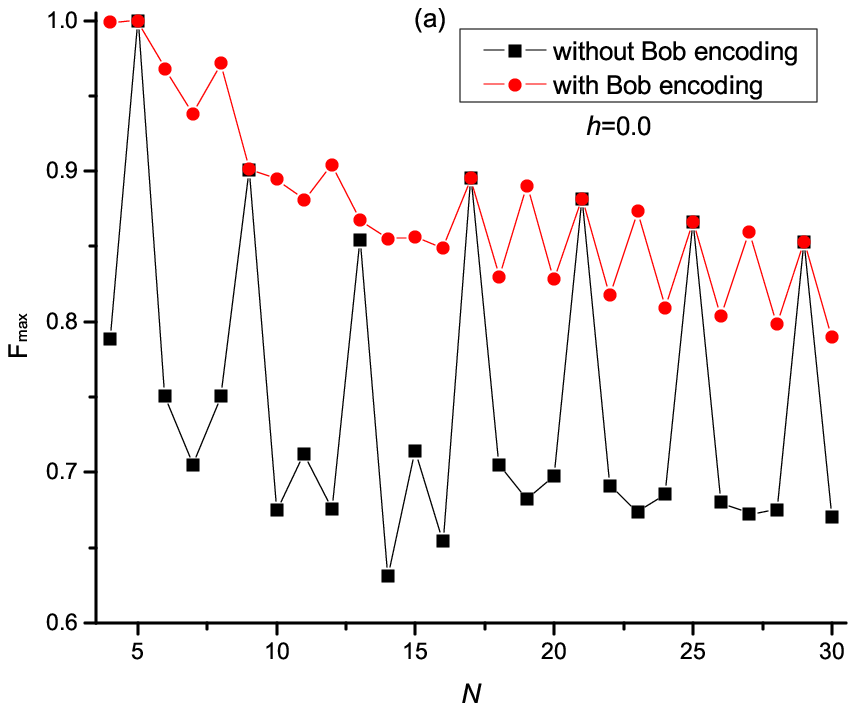} %
\includegraphics[scale=0.6,angle=0]{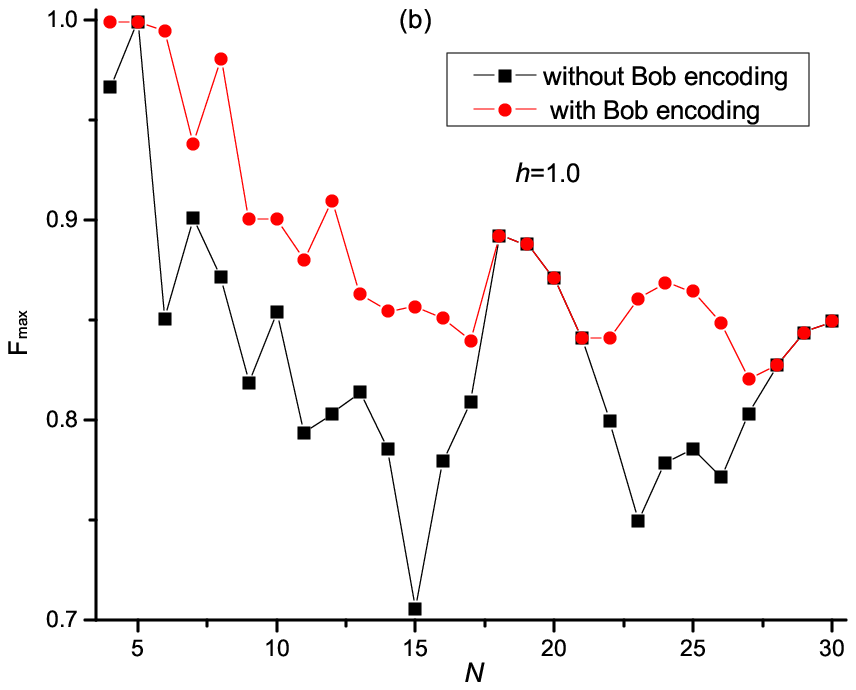}
\par
\caption{(Color online.) The maximum fidelity $F_{max}$ when transferring the quantum state $%
1/2\left\vert 0\right\rangle+\sqrt3/2\left\vert 1\right\rangle$ from
Alice's side to Bob's side as a function of $N$. We search for a maximum in the 
interval t$\in[10,50]$ and set (a) $h=0.0$, and (b) $h=1.0$.}
\par
\label{fig:4}
\end{figure}

In the analysis above, we have only considered a $N=10$ site chain.
We now study the length dependence of the maximal fidelity. In Fig.~4
we compare the success of Alice's state transfer with and without 
Bob's encoding for various chain lengths. In the figures, the horizontal axes represent the number of sites
$N$ and we have selected Alice's state to be $1/2 \left\vert 0\right\rangle + \sqrt{3}/2 \left\vert 1\right\rangle$ in 
both figures. 
For a given $N$, the maximum fidelity $F_{\max }$ is determined numerically in a range $%
\theta _{2}\in \lbrack 0,\pi ]$ and $\phi _{2}\in \lbrack 0,2\pi ]$. Fig.~4(a) and (b) correspond 
to the cases $h=0.0$ and $h=1.0$, respectively. The plots reveal several interesting features. 
First of all, we find that the maximum fidelity of Alice's state transfer is generally enhanced 
when Bob encodes an appropriate state, regardless of the presence or absence of an external field. 
Our numerical results show that when $N=5$, a near
perfect state transfer ($F_{\max }\approx 1$) can be obtained in many different cases. Secondly, when $h=0$ (Fig. 4(a)),
there are particular chain lengths for which the maximum fidelity is independent of Bob's presence, namely 
chains which have $N= 4n+5$ $(n=0,1,2,...)$ sites (except for the slight deviation at $N=13$). This 
property is lost however when an external field is applied. For instance, when $h=1.0$ Alice can 
obtain a higher quality state transfer with Bob's presence for all chain lengths we consider except 
for $N=18, 21$ (see Fig.~4(b)). Finally, for
any practical communication protocol it is important to know the
time $\tau$ at which the fidelity gains its maximum. As an example, for an isolated ($h=0.0$) chain 
consisting of N=10 sites we find that Alice's transmitted state reaches a maximum fidelity 
at time $\tau = 23.1$ when Bob encodes a similar state while $\tau = 29.2$ when Bob 
is not present. When h=1.0,
$\tau = 23.0$ with Bob's encoding, and $\tau = 29.1$ without
Bob's encoding. Thus when Bob encodes an appropriate state Alice can obtain both a higher 
transmission fidelity as well as an increased transfer speed.

\section*{CONCLUSION}

In conclusion, we have investigated the effects of duplex quantum communication 
through an unmodulated spin chain. 
A sophisticated quantum computer will undoubtedly rely on data multiplexing 
so it is an important task to establish potential avenues for two-way 
communication. We have shown that spin chains are indeed viable candidates 
for this purpose. Specifically, we have shown that the transmission fidelity 
at each end of a spin chain can usually be enhanced by the presence of a second party.

Our initiative opens the door for a broad investigation of multi-party communication through 
spin networks and may find applications in many experimental
systems such as quantum dots \cite{Petta2005}, optical
lattices \cite{Simon2011}, or NMR \cite{Zhang2005,Cappellaro2007}.

\section*{ACKNOWLEDGMENTS}

This material is based upon work supported by the National Science
Foundation under Grant No. 11005099 and supported by "the
Fundamental Research Funds for the Central Universities" under Grant
No. 201013037.

\end{document}